\documentstyle[epsf,twocolumn,seceq,office]{jpsj}

\def\runtitle{
Relativistic Band-Structure Calculations for CeTIn$_5$(T=Ir and Co)
}
\def\runauthor{Takahiro {\sc Maehira}, Takashi {\sc Hotta},
Kazuo {\sc Ueda}, and Akira {\sc Hasegawa}}

\title{
Relativistic Band-Structure Calculations for CeTIn$_5$(T=Ir and Co)
and Analysis of the Energy Bands by Using Tight-Binding Method}

\author{Takahiro {\sc Maehira}$^{1}$, Takashi {\sc Hotta}$^{1}$,
Kazuo {\sc Ueda}$^{2,1}$, and Akira {\sc Hasegawa}$^{3}$}

\inst{$^1$Advanced Science Research Center,
Japan Atomic Energy Research Institute,
Tokai, Ibaraki 319-1195 \\
$^2$Institute for Solid State Physics,
University of Tokyo,
Kashiwa, Chiba 277-8581 \\
$^3$Niigata University,
Niigata, Niigata 950-2181}

\recdate{\today}

\abst{
In order to investigate electronic properties of recently discovered
heavy fermion superconductors CeTIn$_5$ (T=Ir and Co),
we employ the relativistic linear augmented-plane-wave (RLAPW) method
to clarify the energy band structures and Fermi surfaces of those
materials.
The obtained energy bands mainly due to the large hybridization
between Ce $4f$ and In 5$p$ states well reproduce the Fermi surfaces
consistent with the de Haas-van Alphen experimental results.
However, when we attempt to understand magnetism and superconductivity
in CeTIn$_5$ from the microscopic viewpoint,
the energy bands obtained in the RLAPW method are too complicated
to analyze the system by further including electron correlations.
Thus, it is necessary to prepare a more simplified model, keeping
correctly the essential characters of the energy bands obtained
in the band-structure calculation.
For the purpose, we construct a tight-binding model for CeTIn$_5$
by including $f$-$f$ and $p$-$p$ hoppings as well as $f$-$p$
hybridization, which are expressed by the Slater-Koster integrals,
determined by the direct comparison with the band-calculation result.
Similarity and difference between CeIrIn$_5$ and CeCoIn$_5$
are discussed based on the obtained tight-binding model,
suggesting a significant importance of the effect of
crystalline electric field to understand the difference in
electronic properties among CeTIn$_5$.
}

\kword{Relativistic linear APW method, CeIrIn$_5$, CeCoIn$_5$,
Fermi surface, tight-binding model}

\begin{document}
\sloppy
\maketitle

%
%
\section{Introduction}

Since the establishment of the density functional theory,\cite{DFT}
the band-structure calculation method has been developed as
a powerful and practical tool to analyze electronic properties
of various kinds of materials from the first-principle viewpoint.
Especially, due to the recent rapid developments in computational
technology, significant improvements have been achieved in the
potential of the band-structure calculation.\cite{review}
For instance, nowadays it is possible to treat the complicated
system including one hundred atoms in the unit cell.

On the other hand, in order to include electron correlation effects
beyond the Hartree-Fock approximation, much efforts have been made
to improve the local-density approximation (LDA).
Along this direction, there have been steady improvements in
approximations, such as the generalized gradient approximation
\cite{GGA} and LDA+U method.\cite{LDA+U}
However, since this is essentially a trial to include the many-body
effect with high accuracy, it is difficult to obtain rapid improvements
along this direction,
although it is a challenging and crucial problem in the research
field of condensed matter theory.
In addition, even if such an improved approximation has been
established in many-body physics,
there still remains another important step to incorporate it into
the band-structure calculation.
Although this difficulty is believed to be overcome in future,
at least at present, it becomes a hard task to have significant
developments immediately.

However, numbers of target materials of the band-structure
calculation has been rapidly increased.
Typically, strongly correlated electron systems
such as transition metal oxides and $f$-electron compounds
have attracted much attentions in recent decades.
In the research field of strongly correlated electron systems,
novel magnetism and unconventional superconductivity have been
the central topics both from experimental and theoretical sides.
In order to understand those intriguing phenomena,
it is essentially important to include correlation effects
based on the concrete knowledge on electronic properties.

For the purpose, it is necessary to promote a couple of 
theoretical researches in parallel with different viewpoints.
Namely, one research is to analyze precisely the energy-band structure
and Fermi surfaces by using the appropriate band-structure calculation
techniques, in order to obtain correct information about
the electronic properties around the Fermi energy,
which can be directly compared with the result of the
angle resolved photoemission experiment.
Another is to construct a simplified tight-binding model
to reproduce well the energy-band structure around the Fermi energy,
in order to include the effect of electron correlation
in this simplified model.
We believe that those two types of researches should be complementary
to each other in order to make significant progress in our understandings on
novel magnetism and unconventional superconductivity,
although it is quite important to develop methods and techniques in
each research.


As is well known, the above framework works quite well
in $3d$-electron materials.
For instance, in high-$T_{\rm c}$ cuprates,
the result of the band-structure calculation based on the LDA
can be reproduced well by using the tight-binding model.
Further adding short-range Coulomb interactions to this
tight-binding model, we can obtain the so-called
Hubbard Hamiltonian, which has been widely accepted
as a canonical model for $3d$ electron systems.
Based on the Hubbard Hamiltonian,
it has been successful to understand several
anomalous properties of high-$T_{\rm c}$ cuprates.
Note that this success is not due to the speciality of $3d$ electron.
Also in the $4d$ electron system such as Sr$_2$RuO$_4$,
which is an exotic material exhibiting triplet superconductivity,
\cite{ruthenate}
three-band Hubbard Hamiltonian can reproduce well the Fermi surface
structure and it becomes an appropriate microscopic model
to investigate magnetism and superconductivity in ruthenates.
\cite{Takimoto}

When we turn our attentions to superconductivity
in $f$-electron systems, unfortunately, our understandings have
been almost limited in the phenomenological level.
One reason is that in $f$-electron compounds,
the band-structure calculation method itself should be improved,
since the relativistic effect cannot be simply neglected in those materials.
In general, the Coulomb interaction between nucleus and electrons
in proportion to the atomic number
becomes very strong in ions including $f$ electrons, indicating that
the velocity of electron included in lanthanide and actinide ions
easily becomes a fraction of light velocity.
A simple way to take into account such a relativistic effect
is to treat the spin-orbit interaction as a perturbation
in the non-relativistic band-structure calculation method,
leading to a handy method to grasp the feature of some $f$-electron
compounds.
However, this is not the first-principle method to take into account
the relativistic effect, since the magnitude of spin-orbit interaction
should be adjusted by hand.
This point has been improved by the {\it relativistic band-structure
calculation technique}, where the Dirac equations, not the
non-relativistic Schr\"odinger equations, are directly solved
to determined the one-electron state.\cite{Loucks}
After the improvements by Hasegawa and co-workers,\cite{Hase1,HaseYama,Higu1}
it has been possible to obtain the reliable electronic band-structure
and reproduce correctly the de Haas-van Alphen (dHvA) 
experimental results for $f$-electron systems.


Due to the developments in the relativistic band-structure
calculation techniques,
we seem to recover the situation quite similar to
that of $d$-electron systems, namely, the tight-binding model can be
constructed due to the comparison with the relativistic band-structure
calculations for $f$-electron systems.
However, we should note that due to the large spin-orbit coupling,
the meaning of one-electron state becomes complicated
in $f$-electron systems, in which the total angular momentum $j$,
neither spin nor orbital degree of freedom,
is the only good quantum number to specify the one-electron state
at each site.
Thus, it is still a non-trivial problem to construct the
tight-binding model for $f$-electron systems,
even if the relativistic band-structure calculation results are at hand. 
In this paper, we attempt to show that it is possible to construct
the tight-binding model
for $f$-electron systems by using the basis to diagonalize the $z$-component
of total angular momentum $j$,
so as to reproduce the result of the relativistic band-structure calculation.
Then, it is suggested that such a model should be useful for the further 
microscopic discussion on magnetism and superconductivity in $f$-electron
compounds.


As a typical example for our trial to construct the model,
in this paper we focus on recently discovered Ce-based superconducting
compounds CeTIn$_5$ (T=Rh, Ir, and Co),\cite{Ce115-1,Ce115-2,Ce115-3}
which are frequently referred to as ``Ce-115'' materials.
A surprising point is that CeCoIn$_5$ exhibits the superconducting
transition temperature $T_{\rm c}$=2.3K,
which is the highest among yet observed for heavy fermion materials
at ambient pressure.
On the other hand, CeIrIn$_5$ shows $T_{\rm c}$=0.4K
which is much less than that of CeCoIn$_5$.
Note that CeRhIn$_5$ is antiferromagnet with the N\'eel temperature
$T_{\rm N}$=3.8K at ambient pressure, while under high-pressure
it becomes superconducting with $T_{\rm c}$=2.1K.
It is not well understood what is the key issue to
characterize the difference in the ground state
of these Ce-115 compounds.
In order to clarify this point, it is necessary to establish
a simplified model to include electron correlation effects from
the microscopic viewpoint.

Among Ce-115 compounds, the dHvA effect has been successfully observed
in CeIrIn$_5$ and CeCoIn$_5$,\cite{Haga,Settai}
both of which have the huge electronic specific heat coefficient
as large as several hundreds mJ/K$^{2}\cdot$mol.
The angular dependence of major experimental dHvA frequency branches is
well explained by a quasi two-dimensional Fermi surface,\cite{Haga,Settai}
which is a clear advantage when we construct a model Hamiltonian,
since we can restrict ourselves in the two-dimensional case.
This is another reason why we choose the Ce-115 compound as a
typical example to attempt to construct a tight-binding model.

In this paper, first we employ the relativistic band-structure
calculation method for CeTIn$_5$ (T=Ir and Co) to investigate
the electronic properties in detail.
It is found that there exists only a small difference between
CeIrIn$_5$ and CeCoIn$_5$ in the energy bands around the Fermi level
and the structure of Fermi surfaces.
Namely, within the band-structure calculations, those two compounds
cannot be distinguished, although $T_{\rm c}$'s are quite different.
Such a difference should originate from the issue with small energy-scale
which cannot be included in the band-structure calculations.
Then, in order to investigate such a small energy scale feature,
we construct a tight-binding model including Ce $4f$ and In $5p$ electrons,
and determine several parameters in the model
by comparing with the band-structure calculation results.
We reanalyze the electronic properties of CeTIn$_5$ to clarify that
the effect of crystalline electric field (CEF) is a key issue to distinguish
CeIrIn$_5$ and CeCoIn$_5$.

The organization of this paper is as follows.
In Sec.~2, we show the relativistic band-structure
calculation results both for CeIrIn$_5$ and CeCoIn$_5$,
to discuss
the obtained energy band structure and the Fermi surfaces.
In Sec.~3, we construct the tight-binding
model for CeTIn$_5$ by using the Slater-Koster integrals.
Then, we make comparison between the relativistic band-structure
calculation results and the dispersion obtained from our tight binding
model.
Finally in Sec.~4, we summarize this paper.
Throughout this paper we use the energy units as $k_{\rm B}$=$\hbar$=1.

\begin{figure}[t]
\label{fig:crys}
\centerline{\epsfxsize=6.5truecm \epsfbox{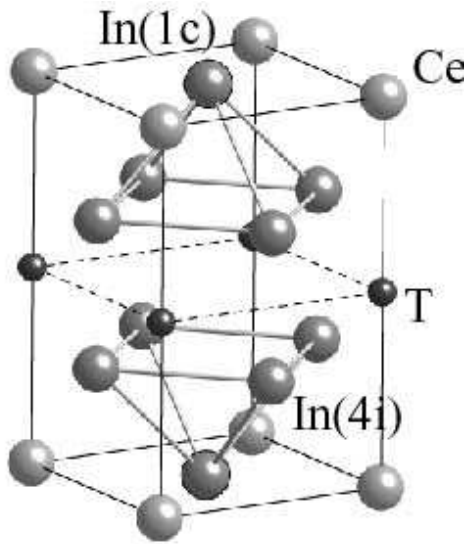} }
\caption{Crystal structure of ${\rm CeTIn_{5}}$, 
called the ${\rm HoCoGa_{5}}$-type tetragonal structure.}
\end{figure}
\begin{table}[t]
\caption{Lattice constants and atomic positions of CeIrIn$_5$ and
CeCoIn$_5$ determined experimentally.~\cite{Haga}
The labels for atoms are referred in Fig.~1.}
\begin{tabular}{lll}
    & CeIrIn$_5$ & CeCoIn$_5$ \\
$a$ & 4.666 \AA (=8.818 a.u.) & 4.612 \AA (=8.714 a.u.) \\
$c$ & 7.516 \AA (=14.205 a.u.) & 7.549 \AA (=14.264 a.u.) \\
Ce  & (0, 0, 0)  & (0, 0, 0) \\
Ir  & (0, 0, 1/2) & (0, 0, 1/2) \\
In($1c$) & (1/2, 1/2, 1/2) & (1/2, 1/2, 1/2) \\
In($4i$) & (0, 1/2, $\pm$0.30528) & (0, 1/2, $\pm$0.305)\\
         & (1/2, 0, $\pm$0.30528) & (1/2, 0, $\pm$0.305) \\
\end{tabular}
\end{table}

%
%
\section{Energy Band Calculations for CeTIn$_5$}

In order to calculate the electronic energy band structure of
cerium compounds, in general, relativistic effects should be seriously
taken into account.\cite{Hase1}
As is well known, in the hydrogen atom, $v$/$c$ is just equal
to the fine structure constant 1/137, which is negligibly small,
where $v$ is the velocity of electron and $c$ is the light velocity.
However, in the atom with the atomic number $Z$,
$v$/$c$ for the electron in the most inner orbital (K-shell)
is given by $Z/137$,
indicating that the velocity of the K-shell electron in the 
uranium atom with $Z$=92 becomes 67\% of the light velocity.
This is simply understood as follows:
Due to the strong Coulomb attraction enhanced by $Z$,
electrons near the heavy nucleus must move with a high speed
in order to keep their stationary motion.
Thus, the relativistic effect on the inner electrons
becomes more significant in an atom with larger $Z$.

In order to take into account major relativistic effects such as
the relativistic energy shifts, the relativistic screening effects,
and the spin-orbit interaction in the energy 
band structure calculation, Loucks has derived 
a relativistic augmented plane wave (APW) method
based on the Dirac one-electron wave equation.\cite{Loucks}
Although his method was a natural extension of Slater's non-relativistic 
APW method to a relativistic theory, it included a couple of problems.
For instance, the symmetrization of the basis 
functions was not taken into account and 
the method was not self-consistent.
These points have been improved by Yamagami and Hasegawa.\cite{HaseYama}
Note that Andersen has first suggested a linealized technique to
include the relativistic effect in the APW method.\cite{Ander}

Among several methods, in this paper we employ a relativistic
linear augmented-plane-wave (RLAPW) method.
The exchange and correlation potential is considered within
LDA, while the spatial shape of the one-electron potential
is determined in the muffin-tin approximation.
The self-consistent calculation is performed by using the
lattice constants which are determined experimentally.

\subsection{Basic properties of CeTIn$_5$}

Ce-115 materials are categolized into 
the ${\rm HoCoGa_{5}}$-type tetragonal structure as
shown in Fig.~1, which is characterized by the space group
${\rm P4/mmm}$ (No.~123) and ${\rm D_{4h}^{1}}$.
Note that one molecule is contained per primitive cell. 
The lattice constants $a$ and $c$ as well as positions
of all atoms in the unit cell are listed in Table I.

The $4f$ electrons in ${\rm CeTIn_{5}}$ (T=Ir and Co)
are assumed to be itinerant.
The iteration process for solving the Dirac one-electron equation 
starts with the crystal charge density that is constructed by 
superposing the relativistic atomic charge densities for neutral 
atoms Ce([Xe]$4f^{1}5d^{1}6s^{2}$), Co([Ar]$3d^{7}4s^{2}$),
Ir([Xe]$4f^{14}5d^{7}6s^{2}$), and In([Kr]$4d^{10}5p^{1}5s^{2}$),
where [Xe], [Ar], and [Kr] symbolically indicate the
closed electronic configuration for xenon, argon, and krypton,
respectively. 
In the calculation for the atoms, the same exchange and correlation 
potential are used as for the crystal. 
We assume that 
the Xe core state except the $5p^{6}$ state for Ce,
the Xe core state into the $4f^{14}$ for Ir,
the Ar core state for Co, and
the Kr core state for In are unchanged during the iteration.
Namely, the frozen-core approximation 
is adopted for these core states in the calculation for the crystal.
In the relativistic atomic calculation, the spin-orbit splitting in the 
Ce $4f$, Ce $5d$, Co $3d$, Ir $5d$, and In $5p$ state are found to be
25 mRyd., 15 mRyd., 14 mRyd., 90 mRyd., and 18 mRyd., respectively.
Note here that mRyd. denotes milli-Rydberg and 1 Ryd.=13.6 eV.

In each iteration step for the self-consistent calculation processes,
a new crystal charge density is constructed using eighteen $k$ points,
which are uniformly distributed in the irreducible 1/16 part of
the Brillouin zone.
At each $k$ in the Brillouin zone, 431 plane waves are adopted 
under the condition $|k+G|$$\leq$$4.0(2\pi/a)$
with $G$ the reciprocal lattice vector, and angular momentum up to
$\ell_{\rm max}$=8 are taken into account.

\subsection{Results for CeIrIn$_5$}

First let us discuss the calculated results for CeIrIn$_5$,
as shown in Fig.~2, in which we depict the energy band
structure along the symmetry axes in the Brillouin zone
in the energy region from $-0.5$Ryd. to 1.0Ryd.
Note here that the three Ce $5p$ and twenty-five In $4d$ bands
in the energy range between $-1.0$Ryd. and $-0.6$Ryd. are not shown
in Fig.~2, since those bands are irrelevant to the present discussion.
The Fermi level $E_{\rm F}$ is located at 0.416 Ryd. and
in the vicinity of $E_{\rm F}$, there occurs a hybridization between
the Ce 4$f$ and In 5$p$ states.
Above $E_{\rm F}$ near M point, the flat 4$f$ bands split into two
groups, corresponding to the total angular momentum $j$=5/2
(lower bands) and 7/2 (upper bands).
The magnitude of the splitting between those groups is estimated
as 0.4 eV, which is almost equal to the spin-orbit splitting
in the atomic $4f$ state.
\begin{figure}[t]
\centerline{\epsfxsize=7.5truecm \epsfbox{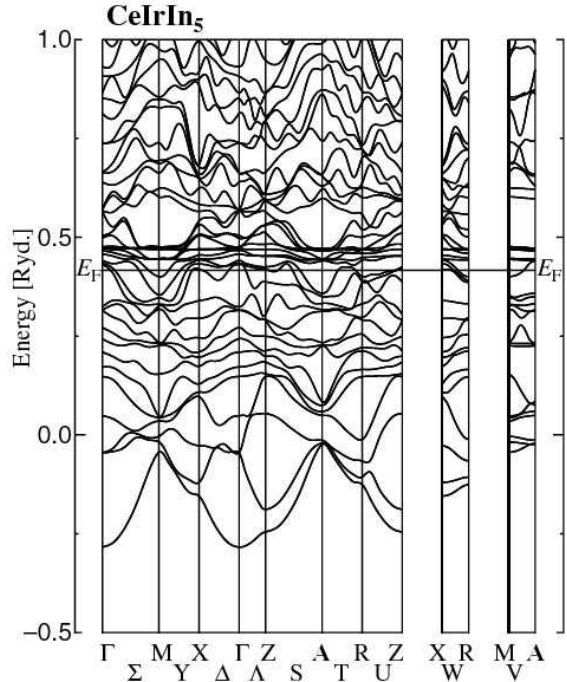} }
\caption{Energy band structure for CeIrIn$_5$ calculated by
using the self-consistent RLAPW method.
$E_{\rm F}$ indicate the position of the Fermi level.}
\end{figure}
\begin{table}[t]
\caption{
The number of valence electrons in the Ce APW sphere,
the Ir APW sphere, and the In APW sphere partitioned 
into angular momentum.}
\begin{tabular}{lrrrr}
\multicolumn{1}{c}{ } & 
\multicolumn{1}{c}{$s$} & 
\multicolumn{1}{c}{$p$} & 
\multicolumn{1}{c}{$d$} & 
\multicolumn{1}{c}{$f$} \\ \hline
Ce & 0.28 & 6.16 & 2.05 & 1.25 \\
Ir & 0.54 & 0.38 & 7.33 & 0.02 \\
In($1c$) & 1.02 & 0.66 & 9.84 & 0.01 \\
In($4i$) & 3.74 & 2.69 & 39.31 & 0.09 \\
\end{tabular}
\end{table}

The number of the valence electrons in the APW sphere is partitioned 
into the angular momentum and listed in Table~II. 
There are 8.79 valence electrons outside the APW sphere
in the primitive cell.
Each Ce APW sphere contains about 1.25 electrons in the $f$ state.
We have obtained similar results for the number of electrons in the 
$f$ state per Ce APW sphere in ${\rm CeSn_{3}}$\cite{Hase2}, 
CeNi\cite{Yama1}, and ${\rm CeRu_{2}}$\cite{Higu}
, although there is difference in the number of electrons in the $d$ state.
We have also found that 0.1-0.2 electrons are contained
in the $f$ state of La APW sphere in corresponding La compounds
and these electrons do not have an atomic 
character but a plane-wave character.
This means that just one $4f$ electron per Ce atom becomes itinerant
in the ground state in ${\rm CeIrIn_{5}}$ and other paramagnetic
Ce compounds.

The total density of states at $E_{\rm F}$ is evaluated as 
$N(E_{\rm F})=134.6$ states/Ryd.cell.
By using this value, the theoretical specific heat coefficient
$\gamma_{\rm band}$ is estimated as 23.3 mJ/K$^2 \cdot$mol.
We note that the experimental electronic specific heat coefficient 
$\gamma_{\rm exp}$ is 750.0 mJ/K$^2 \cdot$mol.
We define the enhancement factor for the electronic specific heat 
coefficient as $\lambda$=$\gamma_{\rm exp}/\gamma_{\rm band}$$-$1,
and in the present case, we obtain $\lambda$=31.2.
The disagreement between $\gamma_{\rm band}$ and $\gamma_{\rm exp}$
values is ascribed to electron correlation effect and
electron-phonon interactions,
which are not fully taken into account in the present
LDA band theory.
\begin{figure}[t]
\centerline{\epsfxsize=8.5truecm \epsfbox{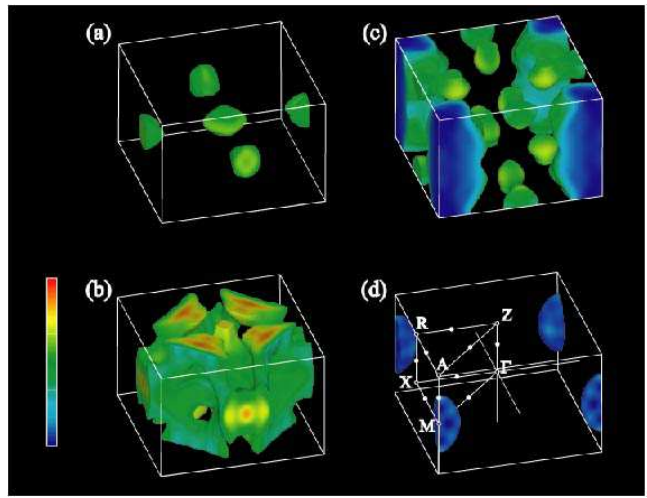} }
\caption{Calculated Fermi surfaces of CeIrIn$_5$ for
(a) 13th band hole sheets, (b) 14th band hole sheets, 
(c) 15th band electron sheets, and 
(d) 16th band electron sheets.
Colors indicate the amount of $4f$ angular momentum character
on each sheet of the Fermi surface.
Red-shift indicate the increase of the admixture of $f$ electrons.
The center of the Brillouin zone is set at the $\Gamma$ point.}
\end{figure}

Now let us discuss the Fermi surfaces of CeIrIn$_5$.
In Fig.~2, the lowest twelve bands are fully occupied.
The next four bands are partially occupied, while higher bands are
empty.
Namely, the 13th, 14th, 15th, and 16th bands crossing
the Fermi level construct the hole or electron sheet of the Fermi surface,
as shown in Fig.~3.
Note that the Fermi surfaces from 13th, 14th, 15th, and 16th bands
are shown in (a), (b), (c), and (d).
The Fermi surface from the 13th band consists of
two equivalent small hole sheets centered 
at the X points and one hole sheet centered at the $\Gamma$ point.
The 14th band constructs a large hole sheet centered at the $\Gamma$
point, which exhibits a complex network consisting of big ``arms''
which lie along the edges of Brillouin zone, as observed in 
Fig.~3(b). 
The 15th band has two kinds of sheets, as shown in Fig.~3(c).
One is a set of two equivalent electron-like pockets,
each of which is centered at the R point. 
Another sheet in the 15th band is a large cylindrical electron sheet
which is centered at the M point.
These electron sheets are characterized by two-dimensional Fermi surfaces.
Namely, they consist of the cylindrical sheet centered at the M point.
The 16th band constructs a small electron sheet,
which is centered at the M point, as shown in Fig.~3(d).
These Fermi surfaces are consistent with the previous results obtained
by the full-potential linealized APW method,\cite{Haga}
but here we point out a possible improvement on our relativistic band-structure
calculation in the appearance of the Fermi surface shown in Fig.~3(d),
which has not been obtained in the previous band-structure
calculations.
The existence of the dHvA branch corresponding to this small Fermi surface
does not contradict the present experimental results, but
such a branch has not been yet identified.
The number of carriers contained in these Fermi-surface sheets are 
0.039 holes/cell, 0.624 holes/cell, 0.626 electrons/cell and 0.037 
electrons/cell in the 13th, 14th, 15th and 16th bands, respectively.
The total number of holes is equal to that of electrons,
which represents that ${\rm CeIrIn_{5}}$ is a compensated metal.

Let us turn our attention to the analysis of the angular momentum
character of the states forming various sheets of the Fermi surface.
Here the most important quantity is the amount of Ce $4f$ character,
which is the partial density of states for Ce $4f$ state
for each point in $k$-space on the Fermi surface. 
This quantity is visualized in Fig.~3,
where the admixture of the Ce $4f$ states is increased as the red shift
in color, as shown in the scale diagram. 
The broad variation of the color from blue to red reflects a 
substantial variation of the $4f$ contribution for different groups 
of states which change in the range from about 5\% to about 85\%. 
However, this distribution is different from part to part
on the Fermi surface.
\begin{figure}[t]
\centerline{\epsfxsize=7.5truecm \epsfbox{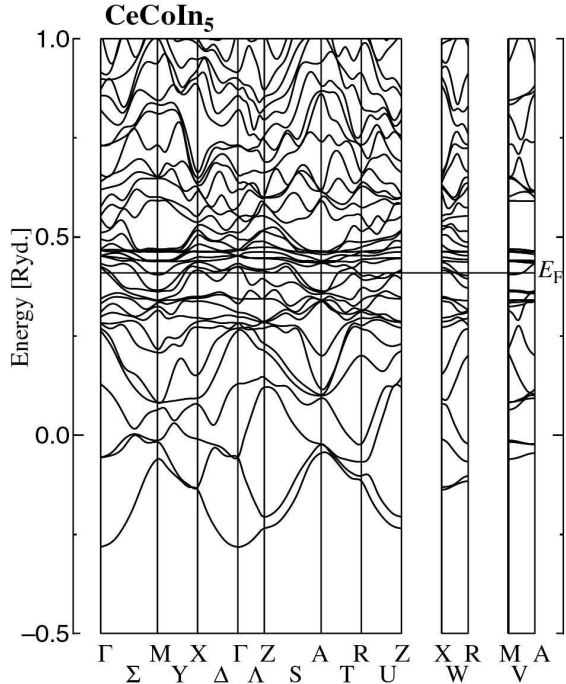} }
\caption{Energy band structure calculated for CeCoIn$_5$ with
the self-consistent RLAPW method.}
\end{figure}
\begin{table}[t]
\caption{
The number of the valence electrons in the Ce APW sphere,
the Co APW sphere, and the In APW sphere partitioned 
into angular momentum.}
\begin{tabular}{lrrrr}
\multicolumn{1}{c}{ } & 
\multicolumn{1}{c}{$s$} & 
\multicolumn{1}{c}{$p$} & 
\multicolumn{1}{c}{$d$} & 
\multicolumn{1}{c}{$f$}   \\ \hline
Ce & 0.29 & 6.16 & 2.11 & 1.28 \\
Co & 0.50 & 0.51 & 7.57 & 0.02 \\
In($1c$) & 0.99 & 0.60 & 9.82 & 0.01 \\
In($4i$) & 3.70 & 2.70 & 39.23 & 0.07 \\
\end{tabular}
\end{table}

\subsection{Results for CeCoIn$_5$}

In this subsection we will discuss the electronic properties of
CeCoIn$_5$.
In Fig.~4, the energy band structure calculated for CeCoIn$_5$
is shown.
Some remarks to depict this figure are the same as
those in Fig.~2.
First of all, we should note that there is no qualitaive difference
in the energy-band structure between CeIrIn$_5$ and CeCoIn$_5$.
Note, however, that the position of $E_{\rm F}$ is a little bit shifted
as $E_{\rm F}$=0.409 Ryd.

The occupation of the bands is also the same as that in CeIrIn$_5$.
Namely, the first twelve bands are fully occupied.
The next four bands are partially occupied, while higher bands are
empty.
The number of the valence electrons in the APW sphere is
listed for each angular momentum in Table III. 
There are 8.63 valence electrons outside the APW sphere in the
primitive cell and each Ce APW sphere contains about 1.28 electrons
in the $f$ state.
The total density of states is calculated at $E_{\rm F}$ as 
$N(E_{\rm F})=154.9$ states/Ryd.cell, leading to
$\gamma_{\rm band}$=26.8 mJ/K$^2 \cdot$mol.
Note that  
$\gamma_{\rm exp}$=300.0 mJ/K$^2 \cdot$mol for CeCoIn$_5$,
smaller than that of CeIrIn$_5$, leading to $\lambda$=10.2
small compared to that of CeIrIn$_5$.
\begin{figure}[t]
\centerline{\epsfxsize=8.5truecm \epsfbox{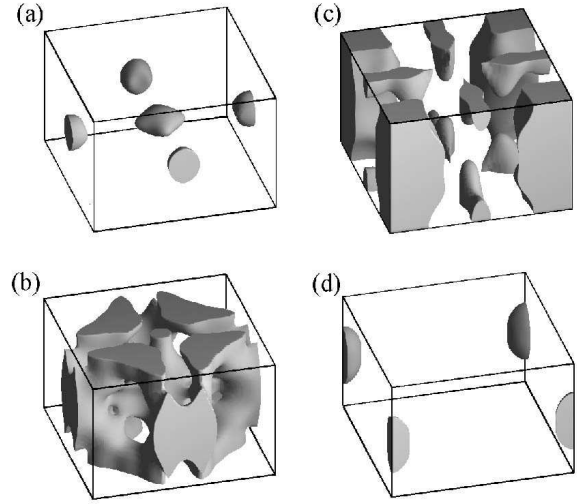} }
\caption{Calculated Fermi surfaces of ${\rm CeCoIn_{5}}$ 
for (a) 13th band hole sheets, (b)14th band hole sheets, 
(c)15th band electron sheets, and 
(d)16th band electron sheets.}
\end{figure}

As for the Fermi surface, the situation is quite similar to CeIrIn$_5$.
Namely, the 13th, 14th, 15th and 16th bands construct the Fermi
surface, as shown in Fig.~5.
When we compare them with those in Fig.~3, it is remarked that
the shapes of the Fermi surfaces are essentially the same as
those of CeIrIn$_5$, although it is possible to point out
difference of Fermi-surface structure in the orbits running
around the bulges of the arms,
which has center at the X point in the 14th band,
and the topology of dumbbell, which has centered at the R point
in the 15th band.
Note also that these Fermi surfaces reproduce well
the previous results.\cite{Settai}
The numbers of carriers contained in these Fermi-surface sheets are
0.032 holes/cell, 0.627 holes/cell, 0.629 electrons/cell, and
0.030 electrons/cell in the 13th, 14th, 15th, and 16th bands,
respectively.
The total number of holes is equal to that of electrons,
indicating that CeCoIn$_5$ is a compensated metal.

The small difference between those two compounds regarding
the band structure around the Fermi level seems to be strange, 
if we recall the experimental facts such as $T_{\rm c}$
and electronic specific heat coefficient $\gamma_{\rm exp}$.
For CeIrIn$_5$, $T_{\rm c}$=0.2K and
$\gamma_{\rm exp}$=750.0 mJ/K$^2 \cdot$mol, 
while for CeCoIn$_5$
$T_{\rm c}$=2.3K and $\gamma_{\rm exp}$=300.0 mJ/K$^2 \cdot$mol.
Namely, even though the band structure is similar to
each other, the physical properties are quite different.
This point is difficult to understand only within the
present band-structrue calculation results.

%
%
\section{Tight-Binding Analysis}

In the previous section we have focussed on the RLAPW band calculation
results, in which the overall framework of 
the band structure around the Fermi energy $E_{\rm F}$ is determined by 
the broad $p$ band and the narrow $f$ bands.
We could obtain the Fermi surfaces consistent with the
experimental results for both materials, although we have found
an unexpected small difference in the energy bands
between CeIrIn$_5$ and CeCoIn$_5$.

In order to shed light on the problem,
we consider the simplified effctive model for CeTIn$_5$
by using the tight-binding method.
For the purpose the two-dimensional lattice composed of
Ce and In ions, are shown in Fig.~6,
since the Fermi surfaces exhibit two-dimensionality,
as has explained in the previous section.
When we consider the tight-binding model on the two-dimensional lattice,
it is enough to take $4f$ orbitals of Ce and $5p$ orbitals of In ions.
In this sense, the model which we will construct is called
the $f$-$p$ model.

\subsection{$f$-$p$ model}

As easily deduced from Fig.~6, the $f$-$p$ model Hamiltonian $H$
should be written as
\begin{equation}
  H = H_{\rm f} + H_{\rm p} + H_{\rm fp}+H_{\rm CEF},
\end{equation}
where the first and second term indicate the hopping of
$f$ and $p$ electrons, respectively, while the third term denotes
the $f$-$p$ hybridization.
The fourth term includes the effect of CEF.
In the following we will consider each term.
Note that in each term of $H$, we simply consider the nearest-neighbor
hoppings of $f$- and $p$-electrons through $\sigma$ bond,
since our purpose here is a construction of a minimal tight-binding
model to discuss magnetism and superconductivity
by further adding Coulomb interactions.

First let us discuss the direct hopping process of $f$-electrons.
Due to the spin-orbit coupling, the energy levels in Ce$^{3+}$ ion
are split into $j$=5/2 sextet and $j$=7/2 octet.
Since the magnitude of this splitting is as large as 0.4 eV,
it is enough to consider only the $j$=5/2 sextet to evaluate
the effective hopping of $f$ electrons.
The sextet in $j$=5/2 are labelled by
$\mu$=$-5/2$, $-3/2$, $\cdots$, $5/2$, with $\mu$ the $z$-component
of $j$, but due to the time reversal symmetry, those states are
classified into three pairs characterized by up and down ``pseudo''
spins.

When we define the second-quantized operator $a_{{\bf i}\mu}$
for the $\mu$-state at site ${\bf i}$, it is convenient to introduce
the new operators $f_{{\bf i}\tau \sigma}$ with pseudospin $\sigma$
as follows:
\begin{equation}
 f_{{\bf i}{\rm a} \uparrow}=a_{{\bf i}-5/2},~
 f_{{\bf i}{\rm a} \downarrow}=a_{{\bf i}5/2},
\end{equation}
for ``a'' orbitals,
\begin{equation}
 f_{{\bf i}{\rm b} \uparrow}=a_{{\bf i} -1/2},~
 f_{{\bf i}{\rm b} \downarrow}=a_{{\bf i} 1/2},
\end{equation}
for ``b'' orbitals,
and 
\begin{equation}
 f_{{\bf i}{\rm c} \uparrow}=a_{{\bf i} 3/2},~
 f_{{\bf i}{\rm c} \downarrow}=a_{{\bf i} -3/2},
\end{equation}
for ``c'' orbitals.
For the standard time reversal operator ${\cal K}$=$-{\rm i}\sigma_yK$,
where $K$ denotes the operator to take complex conjugate,
we can easily show the relation
\begin{equation}
 {\cal K}f_{{\bf i}\tau \sigma}=\sigma f_{{\bf i}\tau -\sigma}.
\end{equation}
Note that this is the same definition for real spin.
\begin{figure}[t]
\centerline{\epsfxsize=7.5truecm \epsfbox{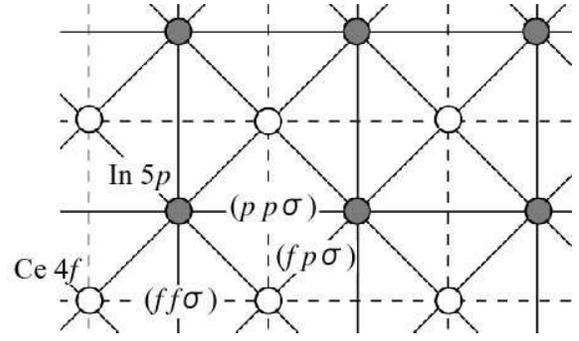} }
\caption{Two-dimensional lattice composed of Ce and In ions.
Open and hatched circles indicate the position of Ce and In ions,
respectively.}
\end{figure}

Now we express the hopping term for $f$ electrons as
\begin{equation}
 H_{\rm f} = \sum_{{\bf i}{\bf a}\tau\tau'\sigma}
 t^{\bf a}_{{\rm f}\tau\tau'} f^{\dagger}_{{\bf i}\tau\sigma}
 f_{{\bf i+a}\tau'\sigma},
\end{equation}
where $t^{\bf a}_{{\rm f}\tau\tau'}$ is the hopping amplitude
of $f$ electron between $\tau$- and $\tau'$-orbitals in
Ce$^{3+}$ ions connected by the vector ${\bf a}$
with ${\bf a}$=${\bf x}$=($\pm$1, 0) and
${\bf a}$=${\bf y}$=(0,$\pm$1).
The hopping amplitude can be expressed as
\begin{equation}
  t_{{\rm f}\tau\tau'}^{\bf x} = {3 \over 56}(ff\sigma)
\left(
\begin{array}{ccc}
5 & -\sqrt{10} & \sqrt{5} \\
-\sqrt{10} & 2 & -\sqrt{2} \\
\sqrt{5} & -\sqrt{2} & 1 \\
\end{array}
\right),
\end{equation}
and
\begin{equation}
  t_{{\rm f}\tau\tau'}^{\bf y} = {3 \over 56}(ff\sigma)
\left(
\begin{array}{ccc}
5 & \sqrt{10} & \sqrt{5} \\
\sqrt{10} & 2 & \sqrt{2} \\
\sqrt{5} & \sqrt{2} & 1 \\
\end{array}
\right),
\end{equation}
where $(ff\sigma)$ is the Slater-Koster integral
between $f$ orbitals through $\sigma$ bond.
\cite{Slater}
Details for the derivation of hopping amplitude
will be discussed elsewhere.\cite{Hotta}

Next let us consider the $p$-electron hopping.
In this case, it is not necessary to consider the
effect of spin-orbit interaction at In site.
Thus, we can consider the hopping of $p$-electrons
with real spin and orbitals.
Furthermore, $p_z$ orbital does not contribute
to the hopping process in the $x$-$y$ plane.
Thus, it is enough to consider the hopping among
$m$=$\pm$1 orbitals ($m$ denotes the $z$-component of
angular momentum $\ell$), since $p_z$ orbital is expressed
only by the $m$=0 state among spherical harmonics for
$\ell$=1.

Note here that it is necessary to redefine spin and orbital
also for $p$ electrons, which should be
consistent with those for $f$ electrons,
since we need to consider the $f$-$p$ mixing later.
When we define the second-quantized operator $c_{{\bf i}m\sigma}$
for the $m$-state with real spin $\sigma$, we can introduce
new operators $p_{{\bf i}\tau \sigma}$ as follows:
\begin{equation}
 p_{{\bf i}{\rm a} \uparrow}=-c_{{\bf i}1 \uparrow},~
 p_{{\bf i}{\rm a} \downarrow}=c_{{\bf i}-1\downarrow},
\end{equation}
for ``a'' orbitals and
\begin{equation}
 p_{{\bf i}{\rm b} \uparrow}=-c_{{\bf i}-1 \uparrow},~
 p_{{\bf i}{\rm b} \downarrow}=c_{{\bf i}1 \downarrow},
\end{equation}
for ``b'' orbitals.
Again we can easily show the relation
\begin{equation}
 {\cal K}p_{{\bf i}\tau \sigma}=\sigma p_{{\bf i}\tau -\sigma}.
\end{equation}
Note that this definition of pseudo spin is consistent with
that of $f$ electron.

After some algebraic calculations, we can obtain
the $p$-electron hopping term as
\begin{equation}
 H_{\rm p} = \sum_{{\bf i}{\bf a}\tau\tau' \sigma}
 t^{\bf a}_{{\rm p}\tau\tau'} p^{\dagger}_{{\bf i}\tau \sigma}
 p_{{\bf i+a}\tau'\sigma},
\end{equation}
Where $t^{\bf a}_{{\rm p}\tau\tau'}$ is the hopping amplitude
of $p$-electron between $\tau$- and $\tau'$-orbitals in
In ions connected by the vector ${\bf a}$,
given as
\begin{equation}
  t_{{\rm p}\tau\tau'}^{\bf x} ={(pp\sigma) \over 2}
\left(
\begin{array}{cc}
1 & -1 \\
-1 & 1 \\
\end{array}
\right),
\end{equation}
and
\begin{equation}
  t_{{\rm p}\tau\tau'}^{\bf y} = {(pp\sigma) \over 2}
\left(
\begin{array}{ccc}
1 & 1 \\
1 & 1  \\
\end{array}
\right),
\end{equation}
where $(pp\sigma)$ is the Slater-Koster integral between $p$ orbitals.

Note here that $p$ electron hopping is expressed by the basis of
spherical harmonics, not the cubic harmonics.
Of course, only for the purpose to show the $p$-electron
hopping, cubic harmonics can provides simpler expression,
since non-zero hopping appears only between $p_x$ ($p_y$)
orbitals along $x$- ($y$-) directions.
However, when we include the $f$-$p$ hybridization,
it is convenient to express it by using the basis of
spherical harmonics.

Now we consider the $f$-$p$ hybridization term.
In the present definitions of pseudo spins for $f$- and $p$-electrons,
it is not necessary to distinguish the real and pseudo spins
in the $f$-$p$ mixing term, if we pay due attentions to
the Clebsch-Gordan coefficients.
Due to the similar calculations to derive $f$-$f$ and $p$-$p$
hopping term, we can obtain the hybridization term as
\begin{equation}
 H_{\rm fp} = \sum_{{\bf i}{\bf b}\tau \tau' \sigma}
 (V^{\bf b}_{\tau \tau'\sigma} f^{\dagger}_{{\bf i} \tau \sigma}
 p_{{\bf i+b}\tau' \sigma} +{\rm h.c.}),
\end{equation}
where the hybridization is given by
\begin{equation}
V^{\bf b}_{\tau \tau'\uparrow} =
V_{\tau \tau'} \equiv
{(fp\sigma) \over 4\sqrt{7}}
\left(
\begin{array}{cc}
\sqrt{15} & i\sqrt{15} \\
i\sqrt{6} & -\sqrt{6} \\
-\sqrt{3} & -i\sqrt{3} \\
\end{array}
\right),
\end{equation}
for ${\bf b}$=$[\pm 1/2,\pm 1/2]$.
Note that ${\bf b}$ is the vector connecting neighbouring
Ce and In ions.
For ${\bf b}$=$[\pm 1/2,\mp 1/2]$,
$V^{\bf b}_{\tau \tau'\uparrow}$=$V_{\tau \tau'}^{*}$.
For down spins, we easily obtain
\begin{equation}
V^{\bf b}_{\tau \tau'\downarrow} =V^{{\bf b}*}_{\tau \tau' \uparrow}.
\end{equation}
By using the operators defined above, we can express
the CEF term as 
\begin{equation}
 H_{\rm CEF} = \sum_{{\bf i}\tau\tau'\sigma}
 B_{\tau\tau'}
 f_{{\bf i}\tau\sigma}^{\dag}f_{{\bf i}\tau'\sigma}
+ \Delta \sum_{{\bf i} \tau\sigma}
 p_{{\bf i}\tau\sigma}^{\dag}p_{{\bf i}\tau\sigma},
\end{equation}
where $B_{\tau\tau'}$ indicates the coefficients
to express the effect of CEF and
$\Delta$ is the energy level for $p$-orbitals
measured from the $f$ electron level.
By consulting with the paper of Hutchings,\cite{CEF}
it is easy to obtain for $j$=5/2 and tetragonal lattice as
\begin{equation}
\begin{array}{rcl}
B_{\rm aa} &=& 10B_2^0+ 60 B_4^0, \\
B_{\rm bb} &=& -8B_2^0+120 B_4^0, \\
B_{\rm cc} &=& -2B_2^0-180 B_4^0, \\
B_{\rm ac} &=& B_{\rm ca}=12\sqrt{5} B_4^4,
\end{array}
\end{equation}
where $B_{\rm p}^{\rm q}$ are the so-called CEF parameters.

All items included in the Hamiltonian have been established, but
it is more convenient to express $H$ in the momentum representation
for the purpose to obtain the energy dispersion.
After straightforward calculations of Fourier transformation,
we finally obtain
\begin{eqnarray}
 H &=& \sum_{{\bf k}\tau\tau'\sigma}
 [(\varepsilon_{{\bf k}\tau\tau'}^{\rm f}+B_{\tau\tau'})
 f^{\dagger}_{{\bf k}\tau \sigma}f_{{\bf k}\tau'\sigma} \nonumber \\
 &+&
 (\varepsilon_{{\bf k}\tau\tau'}^{\rm p}+\Delta \delta_{\tau\tau'})
 p^{\dagger}_{{\bf k}\tau \sigma}p_{{\bf k}\tau'\sigma} \nonumber \\
 &+&
 (V_{{\bf k}\tau \tau' \sigma}
 f^{\dagger}_{{\bf k}\tau \sigma}p_{{\bf k}\tau'\sigma}+{\rm h.c.})],
\end{eqnarray}
where $\delta_{\tau\tau'}$ is the Kronecker delta and
the $f$-electron energy is given by
\begin{equation}
 \varepsilon^{\rm f}_{{\bf k}\tau\tau'} \!=\ {3(ff\sigma) \over 28}
 \left(
 \begin{array}{ccc}
 5\varepsilon_{\bf k} & -\sqrt{10}\gamma_{\bf k} &
 \sqrt{5}\varepsilon_{\bf k} \\
 -\sqrt{10}\gamma_{\bf k} & 2\varepsilon_{\bf k} &
 -\sqrt{2}\gamma_{\bf k} \\
 \sqrt{5}\varepsilon_{\bf k} & -\sqrt{2}\gamma_{\bf k} &
 \varepsilon_{\bf k} \\
 \end{array}
 \right),
\end{equation}
with $\varepsilon_{\bf k}$=$\cos k_x$+$\cos k_y$
and $\gamma_{\bf k}$=$\cos k_x$$-$$\cos k_y$.
The $p$-electron energy is given as
\begin{equation}
 \varepsilon^{\rm p}_{{\bf k}\tau \tau'} \!=\! (pp\sigma)
 \left(
 \begin{array}{cc}
  \varepsilon_{\bf k} & -\gamma_{\bf k} \\
  -\gamma_{\bf k} & \varepsilon_{\bf k} \\
 \end{array}
 \right),
\end{equation}
and the hybridization is written by
\begin{equation}
 V_{{\bf k}\tau \tau' \uparrow} =
 {(fp\sigma) \over \sqrt{7}}
 \left(
 \begin{array}{cc}
  \sqrt{15}c_{\bf k} & i\sqrt{15}s_{\bf k} \\
  i\sqrt{6}s_{\bf k} & -\sqrt{6}c_{\bf k} \\
  -\sqrt{3}c_{\bf k} & -i\sqrt{3}s_{\bf k} \\
 \end{array}
 \right),
\end{equation}
with $c_{\bf k}$=$\cos (k_x/2) \cos (k_y/2)$ and
$s_{\bf k}$=$\sin (k_x/2) \sin (k_y/2)$.
Note that $V_{{\bf k}\tau \tau' \downarrow}$=
$V_{{\bf k}\tau \tau' \uparrow}^*$.

\subsection{Comparison with band calculation results}

In the previous subsection, 
the $f$-$p$ model has been constructed by using
seven parameters $(ff\sigma)$, $(pp\sigma)$, $(fp\sigma)$,
$\Delta$, $B_2^0$, $B_4^0$, and $B_4^4$, which will be
determined in the following conditions:

\noindent
(1) CEF parameters ($B_{2}^{0}$, $B_{4}^{0}$, $B_{4}^{4}$) are
determined from the experimental result for magnetic susceptibility.

\noindent
(2) The positions of the top and bottom of tight-binding bands
correspond to $\Gamma$ point of the 25th and M point of 12th bands,
respectively, in the RLAPW energy band structure.
Note that those are mainly originating from the In 5$p$ state.

\noindent
(3) The energy difference $\Delta$ between $f$ and $p$ levels
is determined to be positive due to the RLAPW band calculation
results.

\noindent
(4) The $f$ electron number per site is fixed as unity, since
Ce$^{3+}$ ion contains one $f$ electron.
However, the total number of electrons is not integer.
It is difficult to reproduce quantitatively the band-structure
calculation results (see, for instance, Table II)
for electron numbers by a two-dimensional situation.

\noindent
(5) We should reproduce the Fermi surfaces constructed
by the 13th, 14th and 15th bands, since they include
significant $f$-electron contribution.

\noindent
(6) Inequalities $(ff\sigma)$$<$$(pp\sigma)$ and
$(ff\sigma)$$<$$(fp\sigma)$ should hold,
since $p$ band is much wider than $f$ band and
$f$-$p$ mixing is large from the RLAPW results.
Note, however, that the relation between $(pp\sigma)$
and $(fp\sigma)$ should be determined by the parameter fitting.
\begin{figure}[t]
\centerline{\epsfxsize=6.5truecm \epsfbox{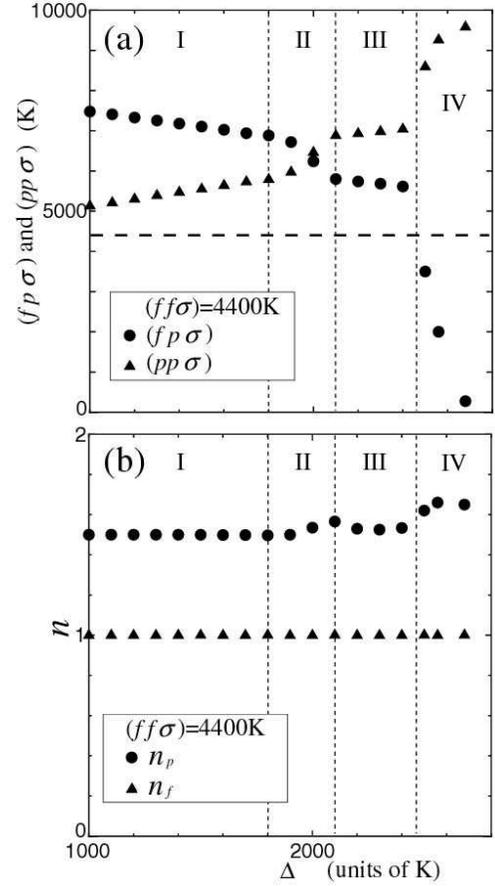} }
\caption{(a) $(fp\sigma)$ and $(pp\sigma)$ as a function of $\Delta$ 
for $(ff\sigma)$=4400K. Note that the horizontal hashed line
indicate the energy of $(ff\sigma)$.
(b) $f$ and $p$ electron number per site as a function of $\Delta$.}
\end{figure}

First let us analyze the energy bands and Fermi surfaces of
CeIrIn$_5$ based on the $f$-$p$ model.
As for CEF parameters, we can determine them as
$B_{2}^{0}$=$-15$K, $B_{4}^{0}$=$0.52$K, and
$B_{4}^{4}$=$0.47$K from the experimental results for
magnetic susceptibility.\cite{Takeuchi}
In Fig.~7(a), we plot $(fp\sigma)$ and $(pp\sigma)$ as a function
of $\Delta$ for $(ff\sigma)$=4400K,
as a typical example to explain the fitting procedure.
Note that we determine the parameters in the two-dimensional tight-binding
model by comparing with the three-dimensional
band-structure calculation results.
Thus, the present $(ff\sigma)$ inevitably becomes large
compared with the actual value in the three-dimensional system.
We can see three regions regarding the number of Fermi surfaces:
One for $\Delta$$<$1800 (region-I),
two for 1800$<$$\Delta$$<$2100 (region-II),
and three for 2100$<$$\Delta$ (region-III and IV).
Due to the condition (5), we need three Fermi surfaces.
Note, however, that for $\Delta$$>$2500 (region-IV),
$(fp\sigma)$ becomes smaller than $(ff\sigma)$,
although we obtain three Fermi surfaces,
which is in contradiction to condition (6).
Thus, we discard the region-IV and choose only region-III,
namely, 2100$<$$\Delta$$<$2500.

Note that $f$-electron number is fixed to unity from the condition (4)
as shown in Fig.~7(b), while $p$ electron number $n_{\rm p}$ is
not fixed.
However, as shown in Fig.~8(b), $n_{\rm p}$ is almost 1.5,
irrespective of $\Delta$.
If we consult with the band-structure calculation results,
as shown in Table I, $n_{\rm p}$=0.66, since we consider the
two-dimensional plane consisting of Ce and In(1c) ions.
Although the band-structure calculation has been done
in the three dimensional situation,
the $f$-$p$ model is constructed by assuming pure two dimensions.
Thus, the discrepancy in $n_{\rm p}$ between the band-structure calculation
and tight-binding model is not primary problem in the following.
It may be possible to consider that the present $f$-$p$ model is
contacted with the ``reservoir'' of $p$ electrons, namely, In(4i) sites,
indicating that $n_{\rm p}$ should not be determined
only within the simple two-dimensional tight-binding model.

After depicting the same figures as Fig.~7(a) for several
values of $(ff\sigma)$, we can determine some region
in parameter space to satisfy the conditions (1)-(6).
In Fig.~8(a), we show a typical result for ($ff\sigma$)=4400K
and $\Delta$=2300K with $(pp\sigma)$=6980K and ($fp\sigma$)=5685K.
The Fermi level is located at the bottom of the Ce 4$f$ bands.
The states near $E_{\rm F}$ have strong Ce $4f$ character,
although they include some In $5p$ admixture.
There are three bands crossing $E_{\rm F}$ along the $\Gamma$-M
lines.
Among them two cross also $E_{\rm F}$ along the X-M direction,
leading to two large cylindrical sheets of Fermi surface.
As shown by solid curves in Fig.~8(b), the Fermi surface of the
tight-binding model from the first band consists of one hole sheet
centered at the $\Gamma$ point.
The second band constructs a cylindrical electron sheet centered
at the M point, as shown by the solid curves in Fig.~8(c).
There exist tiny Fermi surfaces around X-point, although
such Fermi surfaces in the RLAPW results shown by dotted curves
in Fig.~8(b), not in Fig.~8(c).
This disagreement may be due to the simplification of the $f$-$p$ model,
but such tiny Fermi surfaces are not important in the following analysis.
The third band constructs a small cylindrical electron sheet centered
at the M point, as shown in Fig.~8(d).
In total, the overall feature of energy bands around the Fermi level
are well reproduced by the simplified $f$-$p$ model.
\begin{figure}[t]
\centerline{\epsfxsize=8.5truecm \epsfbox{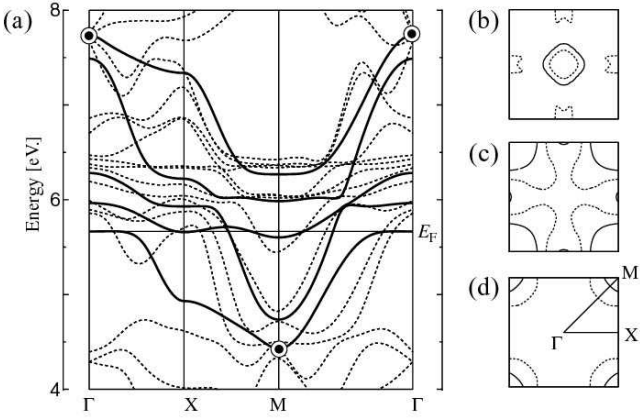}}
\caption{(a)Energy band structures around $E_{\rm F}$ for
CeIrIn$_5$.
The solid and dashed curves indicate the results for tight-binding
and the RLAPW calculation, respectively.
Adjusted RLAPW energy levels are denoted by the circled dots.
Fermi-surface lines discussed here are 
(b)13th band hole sheets, (c)14th band hole sheets, and
(d)15th band electron sheets.
In (b)-(d), solid and dotted curves denote the Fermi-surface lines
for the tight-binding model and the RLAPW results, respectively.}
\end{figure}

Now we turn our attention to CeCoIn$_5$.
CEF parameters of this compound have been determined
experimentally as $B_{2}^{0}$=$-6.8$K,
$B_{4}^{0}$=$0.05$K,
and $B_{4}^{4}$=$2.5$K.\cite{Shishido}
By using these CEF parameters, we have attempted to
reproduce the band-structure calculations, and
finally we can obtain Fig.~9
for $\Delta$=2300K, ($pp\sigma$)=5730K, ($fp\sigma$)=5360K,
and ($ff\sigma$)=4400K.
Again we see that major features of the energy bands of CeCoIn$_5$ 
are basically similar to those of CeIrIn$_5$.
The behavior of the hybridization of energy bands and
the relative position of the $4f$-bands to the other valence bands
are not changed when we compare with those of CeIrIn$_5$.
In fact, the obtained Fermi surfaces are essentially the same
as those of CeIrIn$_5$, as easily understood from the comparison
between Figs.~8 and 9.
\begin{figure}[t]
\centerline{\epsfxsize=8.5truecm \epsfbox{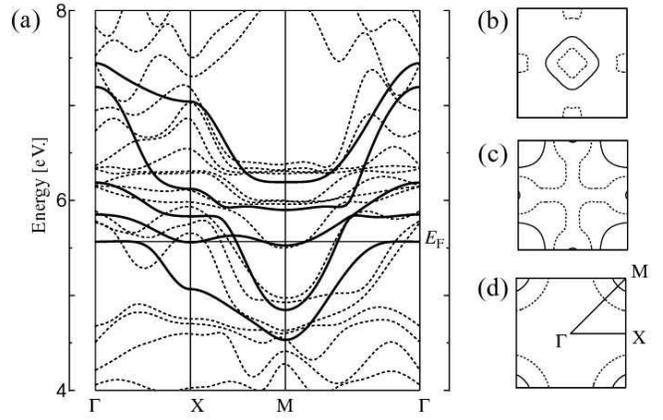}}
\caption{(a)Energy band structures around $E_{\rm F}$ for
CeCoIn$_5$.
Meanings of all notations are the same as those in Fig.~8.
Fermi-surface lines shown here are
(b)13th band hole sheets, (c)14th band hole sheets, and
(d)15th band electron sheets.}
\end{figure}

Here let us consider the meaning of similarity
between those two compounds.
In the $f$-$p$ model we have seven parameters,
which are classified into two groups, i.e., high- and low-energy
groups.
The high-energy group includes ($ff\sigma$), ($pp\sigma$),
($fp\sigma$), and $\Delta$, while low-energy group
consists of CEF parameters, $B_{2}^{0}$, $B_{4}^{0}$, and $B_{4}^{4}$.
Note here that the typical energy scale is 1000K for the former,
while at most 10K for the latter group.
The energy band structure with the RLAPW method
around the Fermi energy $E_{\rm F}$ is mainly determined by 
high-energy group parameters.
On the other hand, the band structure is $not$ sensitive to
CEF parameters, since those are much smaller than
the typical energy scale considered
in the band-structure calculations.
In fact, the resolution of the energy band structure with
the RLAPW method is less than the energy scale of CEF parameters.
Thus, the smallness of difference between CeIrIn$_5$ and
CeCoIn$_5$ suggests that CEF parameters are important to characterize
the difference in Ce-115 materials.
This is consistent with another experimental analysis
for magnetic excitations.\cite{Pagliuso}
\begin{figure}[t]
\centerline{\epsfxsize=6.5truecm \epsfbox{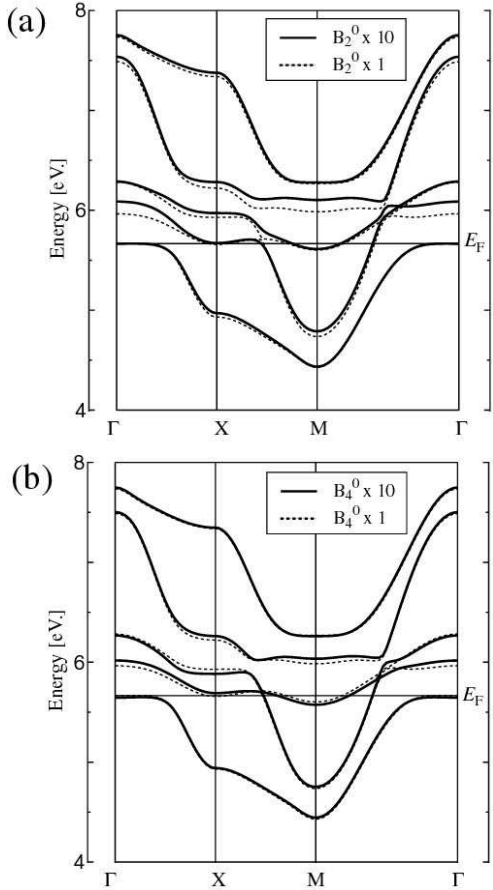}}
\caption{Energy band structures of the $f$-$p$ model
around $E_{\rm F}$ for CeIrIn$_5$
when we increase (a) only $B_2^0$ and (b) only $B_4^0$,
by keeping other parameters invatiant.
The Fermi level for the band-structure for 10$B_{2}^{0}$ is
fixed as that for $B_{2}^{0}$, and $f$-electron number is
unity from the condition (4).}
\end{figure}

\subsection{Role of CEF parameters in Ce-115 compounds}

Since we have confirmed that the overall band-structure is well
reproduced by the tight-binding $f$-$p$ model,
let us now discuss the role played by the CEF parameters in Ce-115
materials based on the $f$-$p$ model from the qualitative viewpoint.
Note again that the effect of CEF parameters cannot be discussed
in the RLAPW results, since it is included only partially
in the band-structure calculations.
In order to see the effect of CEF parameters on the band-structure,
we consider two types of modification in CEF parameters
for CeIrIn$_5$:
In Fig.~10(a), $B_2^0$ is just increased as 10$B_2^0$
by keeping others invariant,
while only $B_4^0$ is increased as 10$B_4^0$ in Fig.~10(b).
As for $B_4^4$, we show no figure, since the band-structure
is not changed significantly, even if we change it as 100$B_4^4$.
Namely, $B_4^4$ is the most insensitive quantity among three
CEF parameters, and thus, we do not discuss the effect of $B_4^4$
furthermore.

In Figs.~10(a) and (b), it is noted that there is clear difference
in the tendency of the change in the band-structure when we 
increase $B_2^0$ or $B_4^0$.
Here note that the solid curves indicates the results for
increasing CEF parameter.
When we increase $B_{2}^{0}$, the energy bands are totally shifted
upward, as observed in Fig.~10(a).
On the other hand, for the increase of $B_4^0$, 
the distance between two bands forming the Fermi surfaces tend to
be narrow, as shown in Fig.~10(b).
Here we focus on two bands forming the main Fermi surfaces,
second and third solid curves in Fig.~8 from the bottom,
since those Fermi surfaces with large volume in the Brillouin zone
includes significant amount of $f$ electrons.
Note that this narrowing effect due to the increase of $B_4^0$
occurs only in bands around the Fermi level, although the shift due to
the change of $B_2^0$ seems to occur in all bands.

It is considered that the physical quantities are sensitive to $B_4^0$,
since those are determined by the band-structure around the Fermi level.
For instance, due to the narrowing effect of effective band-width around
the Fermi level, $N(E_{\rm F})$ (the density of states at the Fermi level)
should be increased with increasing $B_4^0$.
Here we recall the experimetal results on CEF parameters such as
$B_{4}^{0}$=$0.52$K for ${\rm CeIrIn_{5}}$ and 
$B_{4}^{0}$=$0.05$K for ${\rm CeCoIn_{5}}$.
Due to the present discussion, we can conclude that $N(E_{\rm F})$
of ${\rm CeIrIn_{5}}$ becomes larger than
that of ${\rm CeCoIn_{5}}$, if we correctly include the effect of CEF
in the band-structure calculations.
In fact, we have observed that
$\gamma_{\rm exp}$=750.0 mJ/K$^2 \cdot$mol and 300.0 mJ/K$^2 \cdot$mol
for CeIrIn$_5$ and  CeCoIn$_5$, respectively.
We emphasize that the difference in $\gamma_{\rm exp}$ can be understood,
at least qualitatively, by taking into account the effect of
CEF parameters into the $f$-$p$ model.
It is a clear advantage to consider the elecrtonic properties
based on the present tight-binding model, since
the effect of CEF parameters can be included in an explicit manner.

Here it should be also noted that at present, electron correlations
have not been included explicitly.
In general, in actual heavy-fermion materials,
quantities with small energy scale such as CEF parameters
can be significant, since the energy scale of quasi-particles
is much reduced due to the large renormalization effect.
In this sense, it may be considered that our present analysis on the CEF
parameters has been done effectively on the tight-binding model
for $quasi$-$particles$, since we compare it with the band-structure
calculation results, even though correlation effects have not been
satisfactrily included in the band-structure calculations.
In any case, in order to obtain definite conclusion on this issue,
further quantitative works to include electron correlations
are needed in future.

Finally, let us try to consider a possible scenario to understand
the difference in $T_{\rm c}$ between ${\rm CeCoIn_{5}}$
and ${\rm CeIrIn_{5}}$.
It may be risky to conclude something regarding superconductivity
without considering seriously the effect of electron correlations,
since superconductivity in Ce-115 materials has been
believed to be originating from antiferromgantic spin fluctuations.
\cite{Takimoto2}
However, it is an interesting trial to discuss the
effect of $B_4^0$ on $T_{\rm c}$ to clarify the importance of
CEF parameters in Ce-115 materials.
As discussed above, the increase in $B_4^0$ induces the effective narrowing
of the band-width around the Fermi level.
Thus, it is deduced that electron correlation becomes effectively larger
as $B_4^0$ is increased.
In fact, for ${\rm CeRhIn_{5}}$ which is an antiferromagnet
at ambient pressure, $B_4^0$ is estimated as 0.55K,\cite{Takeuchi}
which is slightly larger than that of ${\rm CeIrIn_{5}}$.
This is consistent with the present conclusion.
By following our discussion, it may be possible to conclude that
the magnitude of antiferromgantic spin
fluctuations should be larger in the order of 
${\rm CeCoIn_{5}}$, ${\rm CeIrIn_{5}}$, and ${\rm CeRhIn_{5}}$,
in agreement with the previous discussion on
the effect of level splitting to control
antiferromgantic spin fluctuations.\cite{Takimoto2}
Of course, the present discussion on supercodnuctivity is just in
a qualitative level, but we believe that the effect of CEF should be
one of key issues to distinguish the differece in Ce-115 materials,
in combination with the large renormalization effect due to
electron correlations.

%
%
\section{Summary}

In this paper, we have applied the RLAPW method to the self-consistent
calculation  of the electronic structure for CeIrIn$_5$ and
CeCoIn$_5$ on the basis of the itinerant 4$f$ electron picture.
We have found that a hybridization between the Ce 4$f$ state and
In 5$p$ state occurs in the vicinity of $E_{\rm F}$.
The obtained main Fermi surfaces are composed of two hole sheets and 
one electron sheet, all of which are constructed from the band 
having the Ce 4$f$ state and the In 5$p$ state.

In order to understand the difference in electronic properties
between ${\rm CeIrIn_{5}}$ and ${\rm CeCoIn_{5}}$ even though
the energy-band structure around the Fermi level are quite similar
to each other, we have constructed the $f$-$p$ model based
on the knowledge of the RLAPW band calculation.
It has been found that the size and position of the main 
Fermi surfaces for the $f$-$p$ model agree with the RLAPW ones.
By further analyzing the $f$-$p$ model, we have concluded that
$B_4^0$ among the CEF parameters plays an important role
to determined electronic properties,
consistent with the experimental results oberved in 
${\rm CeIrIn_{5}}$ and ${\rm CeCoIn_{5}}$.

Here it is emphasized that the system can be sensitive to
such small energy quantities as CEF parameters,
since the energy scale of the electron system is much reduced due to
the large renormalization effect in heavy-fermion compound.
Namely, our present conclusion is based on the pre-formation of
heavy quasi-particles.
To compelete the discussion, we need to treat both the effect of CEF and
electron correlation at the same footing on the basis of the microscopic
model, for instance, the $f$-$p$ model with Coulomb interactions.
This point will be discussed elsewhere in future.

In order to make further step to quantitative discussion
on supercondiuctivity,
it is also necessary to analyze the model, in which
short-range Coulomb interaction terms are added to the present
tight-binding model or more simplified $f$-electron
model,\cite{Maehira} leading to a microscopic Hamiltonian
to discuss magnetism and superconductivity of
$f$-electron systems.\cite{Takimoto2}
We believe that in this paper we show one route to arrive at the
elucidation of the mechanism of unconventional superconductivity
in heavy fermion materials.

\section*{Acknowledgements}

We thank M. Higuchi, Y. \=Onuki, P. G. Pagliuso, and T. Takimoto
for discussions.
T. H. and K. U. are supported by the Grant-in-Aid
for Scientific Research from
Japan Society for the Promotion of Science.


\end{document}